\documentclass[12pt]{article}
\usepackage{amsfonts}
\usepackage{amssymb}
\usepackage{amsbsy}
\usepackage{mathrsfs}
\usepackage{amsmath}
\usepackage{graphicx}    
\usepackage{rotating}    
\usepackage{epsfig}
\usepackage{color}

\input epsf.sty

\newlength{\TZ}
\newcommand{\BEQ}{\begin{equation}}     
\newcommand{\BEA}{\begin{eqnarray}}
\newcommand{\BD}{\begin{displaymath}}
\newcommand{\EEQ}{\end{equation}}       
\newcommand{\EEA}{\end{eqnarray}}
\newcommand{\ED}{\end{displaymath}}
\newcommand{\bb}{\begin{eqnarray}}
\newcommand{\ee}{\end{eqnarray}}

\newcommand{\vep}{\varepsilon}          
\newcommand{\vph}{\varphi}              
\newcommand{\vro}{\varrho}              
\newcommand{\D}{{\rm d}}                
\newcommand{\II}{{\rm i}}               
\renewcommand{\Re}{{\rm Re\ }}          
 
\newcommand{\demi}{\frac{1}{2}}         
\newcommand{\wit}[1]{\widetilde{#1}}    
\newcommand{\wht}[1]{\widehat{#1}}      
    
\renewcommand{\vec}[1]{\boldsymbol{#1}} 

\catcode`\@=11
\def\numberbysection{\@addtoreset{equation}{section}
        \def\theequation{\thesection.\arabic{equation}}}

\begin{document}

\begin{titlepage}

\vskip 1.5 cm
\begin{center}
{\Large \bf Non-local meta-conformal invariance \\[0.14truecm] in diffusion-limited erosion}
\end{center}

\vskip 2.0 cm
\centerline{{\bf Malte Henkel}$^{a,b}$}
\vskip 0.5 cm
\begin{center}
$^a$Rechnergest\"utzte Physik der Werkstoffe, Institut f\"ur Baustoffe (IfB), \\ ETH Z\"urich, Stefano-Franscini-Platz 3, 
CH - 8093 Z\"urich, Switzerland\\
$^b$Groupe de Physique Statistique,
D\'epartement de Physique de la Mati\`ere et des Mat\'eriaux,
Institut Jean Lamour (CNRS UMR 7198), Universit\'e de Lorraine Nancy, \\
B.P. 70239,  F -- 54506 Vand{\oe}uvre l\`es Nancy Cedex, France\footnote{permanent address}\\
\end{center}

\begin{abstract}
The non-stationary relaxation and physical ageing in the diffusion-limited erosion process ({\sc dle}) 
is studied through the exact solution of
its Langevin equation, in $d$ spatial dimensions. The dynamical exponent $z=1$, the growth exponent $\beta=\max(0,(1-d)/2)$ and 
the ageing exponents $a=b=d-1$ and $\lambda_C=\lambda_R=d$ are found. 
In $d=1$ spatial dimension, a new representation of the meta-conformal Lie algebra, 
isomorphic to $\mathfrak{sl}(2,\mathbb{R})\oplus\mathfrak{sl}(2,\mathbb{R})$, acts as a dynamical symmetry of the 
noise-averaged {\sc dle} Langevin equation. Its infinitesimal generators are non-local in space. 
The exact form of the full time-space dependence of the two-time response function 
of {\sc dle} is reproduced for $d=1$ from this symmetry. 
The relationship to the terrace-step-kink model of vicinal surfaces is discussed. 
\end{abstract}

\vfill

\end{titlepage}

\setcounter{footnote}{0} 


{\bf 1.} The physics of the growth of interfaces is a paradigmatic example of 
the emergence of non-equilibrium phenomena cooperative phenomena 
\cite{Bara95,Krug97,Corw12,Taeu14,Halp15}. The most common universality classes, such as the Edwards-Wilkinson ({\sc ew}) \cite{Edwa82},
Kardar-Parisi-Zhang ({\sc kpz}) \cite{Kard86}, Wolf-Villain ({\sc wv}) \cite{Wolf90} or Arcetri \cite{Henk15} classes, 
are usually specified in terms of models describing the deposition of
particles on a surface, leading to the formation of a fluctuating height profile $h(t,\vec{r})$ of the interface.
The cooperative nature of the phenomenon is expressed in the long-time Family-Viscek scaling behaviour \cite{Fami85} 
of the interface width 
\BEQ \label{1}
w^2(t;L) := \frac{1}{L^d} \sum_{\vec{r}\in\mathscr{L}} \left\langle \left( h(t,\vec{r})-\overline{h}(t) \right) \right\rangle^2 
= L^{2\alpha} f_w\left(t L^{-z}\right) 
\sim \left\{ \begin{array}{ll} t^{2\beta} & \mbox{\rm ~;~~ if $tL^{-z}\ll 1$} \\
                              L^{2\alpha} & \mbox{\rm ~;~~ if $tL^{-z}\gg 1$} \end{array} \right.
\EEQ
on a hyper-cubic lattice $\mathscr{L}\subset\mathbb{Z}^d$ of $|\mathscr{L}|=L^d$ sites, where $\langle .\rangle$ 
denotes an average over many independent samples and $\overline{h}(t) := L^{-d} \sum_{\vec{r}\in\mathscr{L}} h(t,\vec{r})$ 
is the spatially averaged height. 
Herein, $\alpha$ is the {\em roughness exponent}, $\beta$ the {\em growth exponent} and $z=\alpha/\beta>0$ the {\em dynamical exponent}.
The interface is called {\em rough} if $\beta>0$ and {\em smooth} if $\lim_{t\to\infty} w(t)$ is finite. 
While theoretical studies abound, reliable experimental
results are quite recent. Examples for the {\sc kpz} class 
include turbulent liquid crystals, cell colony growth, collo\"{\i}ds, 
paper combustion, auto-catalytic reaction fronts, thin semiconductor films and sedimentation-electrodispersion, 
see \cite{Take14,Halp15} for recent reviews and \cite{Henk15} for a list of measured values of these exponents. 
More subtle aspects can be studied through the non-equilibrium relaxation, 
analogous to physical ageing e.g. in glasses or simple magnets \cite{Henk10}. Analysis proceeds via the
two-time correlator $C(t,s;\vec{r})$ and the two-time response $R(t,s;\vec{r})$. For sufficiently large lattices (where effectively $L\to\infty$)
one expects, in the long-time scaling limit $t,s\to\infty$ with $y :=t/s>1$ fixed, the scaling behaviour
\BEA
C(t,s;\vec{r}) &:=& \left\langle \left(h(t,\vec{r}) - \left\langle \overline{h}(t)\right\rangle \right)
\left(h(s,\vec{0}) - \left\langle \overline{h}(s)\right\rangle \right) \right\rangle 
\hspace{0.55truecm}\:=\: s^{-b} F_C\left( \frac{t}{s}; \frac{\vec{r}}{s^{1/z}} \right)  \label{1.2} \\
R(t,s;\vec{r}) &:=& \left. \frac{ \delta \left\langle h(t,\vec{r}) - \overline{h}(t)\right\rangle}{\delta j(s,\vec{0})}\right|_{j=0} 
\:=\: \left\langle h(t,\vec{r}) \wit{h}(s,\vec{0}) \right\rangle \:=\: s^{-1-a} F_R\left( \frac{t}{s}; \frac{\vec{r}}{s^{1/z}} \right)
\label{1.3}
\EEA
where spatial translation-invariance has been implicitly admitted and $j$ is an external field 
conjugate\footnote{In the context of Janssen-de Dominicis theory, $\wit{h}$ is the conjugate response field to $h$, see \cite{Taeu14}.}  
to $h$. The {\em autocorrelation exponent} $\lambda_C$ and the {\em autoresponse exponent} $\lambda_R$ are defined from the asymptotics
$F_{C,R}(y,\vec{0}) \sim y^{-\lambda_{C,R}/z}$ as $y\to\infty$. For these non-equilibrium exponents, one has $b=-2\beta$ and the bound
$\lambda_C\geq (d+z b)/2$. For the {\sc ew}, {\sc kpz}, {\sc wv} and Arcetri classes, where the dynamical exponent $z\geq\frac{3}{2}$, 
the values of $a,b,\lambda_C,\lambda_R$ have been determined 
either analytically or in simulations \cite{Krec97,Roet06,Darv09,Chou10,Daqu11,Henk12,Halp14,Odor14,Henk15} or else experimentally \cite{Take12}.  

Here, we shall be interested in a different universality class, namely {\em diffusion-limited erosion} ({\sc dle}) \cite{Krug81}, often also
referred to as {\em Laplacian growth}. We shall first derive the width $w(t)$, the correlator $C(t,s;\vec{r})$ and the response $R(t,s;\vec{r})$ from the 
exact solution of the defining Langevin equation. Then, for $d=1$ spatial dimension, 
we shall construct a new representation of the conformal Lie algebra, in terms
of {\it spatially non-local operators}. We shall show that (i) this representation acts as a dynamical symmetry of the equation of motion 
of {\sc dle} and (ii) that for $d=1$, this dynamical symmetry (which  has $z=1$), predicts the form
of the response $R(t,s;\vec{r})$.  

{\bf 2.} The {\em {\sc dle} process} \cite{Krug81} can be defined as a lattice model by considering the diffusive motion of a 
corrosive particle, which starts initially far away from
the interface. When the particle finally reaches the interface, it erodes a particle from that interface. Repeating this process many times,
an eroding interface forms which is described in terms of a fluctuating height $h(t,\vec{r})$. This leads to 
the Langevin equation for $h(t,\vec{r})$ in {\sc dle}. In Fourier space \cite{Krug81,Krug94}
\BEQ \label{4}
\partial_t \wht{h}(t,\vec{q}) = -\nu |\vec{q}| \wht{h}(t,\vec{q}) + \wht{\jmath}(t,\vec{q}) + \wht{\eta}(t,\vec{q})
\EEQ
including the gaussian white noise\footnote{Below, we shall
refer to (\ref{4}) with $\wht{\eta}=0$ as the {\em deterministic part} of (\ref{4}).} $\wht{\eta}$, with the variance 
$\langle \wht{\eta}(t,\vec{q}) \wht{\eta}(t',\vec{q}')\rangle = 2\nu T \delta(t-t')\delta(\vec{q}+\vec{q}')$ and the constants $\nu,T$ and
an external perturbation $\wht{\jmath}$.  
Several lattice formulations of the model exist \cite{Krug81,Yoon03,Aarao07,Zoia07}. Flat and radial geometries are compared in \cite{Heday14}.
Potential applications of {\sc dle} may include contact lines of a liquid meniscus and crack propagation \cite{Joan84}. 
Remarkably, for $d=1$ space dimension, the Langevin equation (\ref{4}) has been argued \cite{Spoh99} to be related to a system of 
non-interacting fermions, conditioned to an a-typically large flux. Consider the {\em terrace-step-kink model} of a vicinal surface, and interpret 
the steps as the world lines of fermions. Its transfer matrix 
is the matrix exponential of the quantum hamiltonian $H$ of the asymmetric XXZ chain \cite{Spoh99}. 
Use Pauli matrices
$\sigma_{n}^{\pm,z}$, attached to each site $n$, such that the particle number at each site is $\vro_n=\demi\left(1+\sigma_{n}^{z}\right)=0,1$.
On a chain of $N$ sites \cite{Spoh99,Popk11,Kare16}
\BEQ \label{xxz}
H = - \frac{w}{2} \sum_{n=1}^{N} \left[ 2 v \sigma_n^{+}\sigma_{n+1}^{-} + 2 v^{-1} \sigma_{n}^{-}\sigma_{n+1}^{+} 
+\Delta \left( \sigma_{n}^{z}\sigma_{n+1}^{z}-1\right) \right]
\EEQ
where $w=\sqrt{pq\,}\, e^{\mu}$, $v=\sqrt{p/q\,}\,e^{\lambda}$ and 
$\Delta=2\left(\sqrt{p/q}+\sqrt{q/p}\right)e^{-\mu}$. Herein, $p,q$ describe the left/right bias of single-particle hopping and $\lambda,\mu$ are the
grand-canonical parameters conjugate to the current and the mean particle number. 
In the continuum limit, the particle density $\vro_n(t) \to \vro(t,r)=\partial_r h(t,r)$ 
is related to the height $h$ which in turn obeys (\ref{4}), with a {\em gaussian white noise} $\eta$ \cite{Spoh99}. This follows from the
application of the theory of fluctuating hydrodynamics, see \cite{Spoh14,Bert15} for recent reviews. 
The low-energy behaviour of $H$ yields the dynamical exponent $z=1$ \cite{Spoh99,Popk11,Kare16}.
\footnote{Empirically, the bubbles in the price of crude oil display dynamical scaling
of the form (\ref{1}) with $z\approx 1$ \cite{Garc16}.} 
If one conditions the system to an a-typically large current, 
the large-time, large-distance behaviour of (\ref{xxz}) has very recently been
shown \cite{Kare16} (i) to be described by a conformal field-theory with central charge $c=1$ and (ii) 
the time-space scaling behaviour of the stationary structure function has been worked out explicitly, for $\lambda\to\infty$. 
Therefore, one may conjecture that the so simple-looking eq.~(\ref{4}) should furnish an effective continuum
description of the large-time, long-range properties of quite non-trivial systems, such as (\ref{xxz}). 
%
%

The solution of (\ref{4}) reads in momentum space 
\BEQ
\wht{h}(t,\vec{q}) = e^{-\nu|\vec{q}|t}\; \wht{h}(0,\vec{q}) + \int_0^t \!\D\tau\: e^{-\nu|\vec{q}|(t-\tau)} 
\left( \wht{\jmath}(\tau,\vec{q}) +\wht{\eta}(\tau,\vec{q})\right)
\EEQ
In this letter, we focus on the non-equilibrium relaxation of {\sc dle}, starting from an an initially flat interface $h(0,\vec{r})=0$. 
If $\wht{\jmath}(t,\vec{q})=0$, the average interface position remains fixed, thus $\langle\wht{h}(t,\vec{q})\rangle=0$ and 
$\langle h(t,\vec{r})\rangle=0$. The two-time correlator and response are 
\begin{subequations} \label{6}
\begin{align}
\wht{C}(t,s;\vec{q},\vec{q}') &:= \left\langle \wht{h}(t,\vec{q}) \wht{h}(s,\vec{q}') \right\rangle
= \frac{T}{|\vec{q}|} \left[ e^{-\nu|\vec{q}||t-s|} - e^{-\nu|\vec{q}|(t+s)} \right] \delta(\vec{q}+\vec{q}')
\label{6C} \\
\wht{R}(t,s;\vec{q},\vec{q}') &:= \left. \frac{\delta \langle \wht{h}(t,\vec{q})\rangle}{\delta \wht{\jmath}(s,\vec{q}')}\right|_{j=0} 
\hspace{0.4truecm}= \Theta(t-s)\; e^{-\nu|\vec{q}|(t-s)}\, \delta(\vec{q}+\vec{q}')
\label{6R}
\end{align}
\end{subequations}
which becomes in direct space, with ${\cal C}_0 := \pi^{-(d+1)/2}\Gamma((d+1)/2)/\Gamma(d/2)$, and for $d\ne 1$
\begin{subequations} \label{7}
\begin{align}
C(t,s;\vec{r}) &= \frac{T {\cal C}_0}{d-1} 
\left[ \left( \nu^2 (t-s)^2 + r^2 \right)^{-(d-1)/2} - \left( \nu^2 (t+s)^2 + r^2 \right)^{-(d-1)/2} \right]
\label{7C} \\
R(t,s;\vec{r}) &= {\cal C}_0\: \Theta(t-s)\:\nu(t-s) \left(\nu^2 (t-s)^2 + r^2\right)^{-(d+1)/2} 
\label{7R}
\end{align}
\end{subequations}
where the Heaviside function $\Theta$ expresses the causality condition $t>s$. In particular, the interface width 
$w^2(t)=C(t,t;\vec{0})$ is 
(apply a high-momentum cut-off $\Lambda$ for $L\to\infty$, if $d>1$)
\BEQ \label{8}
\!\!\!\!\!w^2(t) = \frac{T {\cal C}_0}{1-d} \left[ \left( 2\nu t\right)^{1-d} - {\cal C}_1(\Lambda) \right] \stackrel{t\to\infty}{\simeq}
\left\{ \begin{array}{ll} \!\!T{\cal C}_0{\cal C}_1(\Lambda)/(d-1)        & \mbox{\rm ;\, if $d>1$} \\
                          \!\!T{\cal C}_0 \ln( 2\nu t)                    & \mbox{\rm ;\, if $d=1$} \\
                          \!\!\left[T{\cal C}_0 (2\nu)^{1-d}/(1-d)\right]\!\cdot t^{1-d} & \mbox{\rm ;\, if $d<1$}
        \end{array} \right.
\EEQ
This shows the upper critical dimension $d^*=1$ of {\sc dle}, such that at late times the interface is smooth for $d>1$ 
and rough for $d\leq 1$ \cite{Krug81}. 
In the long-time {\it stationary} limit $t,s\to\infty$ with the time difference $\tau=t-s$ being kept fixed, one has the fluctuation-dissipation relationship
$\partial C(s+\tau,s;\vec{r})/\partial \tau = - \nu T R(s+\tau,s;\vec{r})$. This was to be expected, since there exist lattice model versions 
in the {\sc dle} class which can be  formulated in terms of an equilibrium system \cite{Yoon03}. Finally, in the long-time {\it scaling} limit
$t,s\to\infty$ with $y:=t/s>1$ being kept fixed, one may read off from (\ref{7},\ref{8}) the exponents
\BEQ
\beta=\alpha=\left\{ \begin{array}{ll}  0      & \mbox{\rm ~~;~ if $d>1$} \\
                                        (1-d)/2 & \mbox{\rm ~~;~ if $d<1$} 
                     \end{array} \right. \;\; , \;\; z=1 \;\;,\;\; a=b=d-1 \;\; , \;\; \lambda_C=\lambda_R =d 
\EEQ
In contrast to the interface width $w(t)$, which shows a logarithmic growth at $d=d^*=1$, logarithms cancel in the two-time correlator $C$ and
response $R$, up to {\it additive} logarithmic corrections to scaling. This is well-established in the physical ageing of magnetic systems \cite{Henk10}.

{\bf 3.} Can one explain the form of the two-time scaling functions of the {\sc dle} in terms of a dynamical symmetry~? Such an approach, based on
extensions of the dynamical scaling $t\mapsto b^z t$ and $\vec{r}\mapsto b \vec{r}$ to a larger set of transformations where $b=b(t,\vec{r})$
becomes effectively time-space-dependent, has been applied and tested in the physical ageing of magnetic systems, quenched either to their critical
temperature $T=T_c>0$ or else to $T<T_c$ (where $z=2$), see \cite{Henk10} for a detailed review. More recently, this was also done 
for the relaxation dynamics in interface growth, namely for the the {\sc ew} class \cite{Roet06} where $z=2$ and the 
$(1+1)D$ {\sc kpz} class \cite{Henk12}, where $z=\frac{3}{2}$. These tests mainly involved the fitting of the auto-response $R(t,s;\vec{0})$ to
the exact solutions or the numerical data. Since in the {\sc dle} class, one has $z=1$, a different set of local time-space
transformations must be sought. It might look tempting to consider conformal invariance \cite{Bela84}, 
well-known from equilibrium critical phenomena, by
simply relabelling one of the spatial directions as `time', since this would give $z=1$. However, as we
shall see, a more precise definition is needed. For notational simplicity, we now restrict to the case of $1+1$ time-space dimensions, labelled by
a `time coordinate' $t$ and a `space coordinate' $r$. 

\noindent
{\bf Definition.} {\it 1. A set of {\em ortho-conformal transformations}\footnote{From the greek prefix $o\vro\theta o$: right, standard.} 
(usually called `conformal transformation') $\mathscr{O}$ is a set of maps
$(t,r)\mapsto (t',r')=\mathscr{O}(t,r)$ of local coordinate transformations, 
depending analytically on several parameters, such that angles in the coordinate space of the
points $(t,r)$ are kept invariant. The maximal finite-dimensional Lie sub-algebra of ortho-conformal transformations 
is isomorphic to $\mathfrak{conf}(2)\cong\mathfrak{sl}(2,\mathbb{R})\oplus\mathfrak{sl}(2,\mathbb{R})$. A physical system is 
{\em ortho-conformally invariant} if its $n$-point functions transform covariantly under ortho-conformal transformations. \\
2. A set of {\em meta-conformal transformations}\footnote{From the greek prefix $\mu\vep\tau\alpha$: of secondary rank.}  
$\mathscr{M}$ is a set of maps $(t,r)\mapsto (t',r')=\mathscr{M}(t,r)$, depending analytically on several parameters, whose maximal 
finite-dimensional Lie sub-algebra of meta-conformal transformations is isomorphic to $\mathfrak{conf}(2)$. A physical system is 
{\em meta-conformally invariant} if its $n$-point functions transform covariantly under meta-conformal transformations.}

Hence, ortho-conformal transformations are  also meta-conformal transformations. 

In $(1+1)D$, ortho-conformal transformations are all analytic
or anti-analytic maps, $z\mapsto f(z)$ or $\bar{z}\mapsto \bar{f}(\bar{z})$, of the complex variables $z =t+\II r$, $\bar{z}=t-\II r$.
For our purposes, we restrict here to the projective conformal transformations $z\mapsto \frac{\alpha z+\beta}{\gamma z+\delta}$ 
with $\alpha\delta-\beta\gamma=1$ and similarly for $\bar{z}$. Then the Lie algebra generators $\ell_n = -z^{n+1}\partial_z$ and 
$\bar{\ell}_n = -\bar{z}^{n+1}\partial_{\bar{z}}$ with $n=\pm 1,0$ span the Lie algebra 
$\mathfrak{conf}(2)\cong\mathfrak{sl}(2,\mathbb{R})\oplus\mathfrak{sl}(2,\mathbb{R})$. 
We shall use below the basis\footnote{Interpretation: $X_{-1},Y_{-1}$ generate time- and space-translations, $X_0$ global dilatations 
$t\mapsto bt$, $r\mapsto b r$, $Y_0$ rigid time-space rotations and $X_1,Y_1$ generate the `special' conformal transformations.} 
$X_n := \ell_n +\bar{\ell}_n$ and $Y_n := \ell_n -\bar{\ell}_n$. 
In an ortho-conformally invariant physical system, these generators act on physical `quasi-primary' \cite{Bela84}
scaling operators $\phi=\phi(z,\bar{z})=\vph(t,r)$ and then contain also terms which describe 
how these quasi-primary operators should transform, namely
\BEQ
\ell_n = - z^{n+1}\partial_z - \Delta (n+1) z^n \;\; , \;\; \bar{\ell}_n = -\bar{z}^{n+1}\partial_{\bar{z}} -\overline{\Delta} (n+1) \bar{z}^n
\EEQ
where $\Delta,\overline{\Delta}$ are the conformal weights of the scaling operator $\phi$. Laplace's equation
${\cal S}\phi=4\partial_z \partial_{\bar{z}}\phi=0$ is a simple example of an ortho-conformally invariant system, since the commutator
\BEQ
\left[ {\cal S}, \ell_n \right]\phi(z,\bar{z}) = -(n+1) z^n {\cal S} \phi(z,\bar{z}) - 4\Delta n(n+1) z^{n-1} \partial_{\bar{z}}\phi(z,\bar{z})
\EEQ
shows that for a scaling operator $\phi$ with $\Delta=\overline{\Delta}=0$, 
the space of solutions of the Laplace equation ${\cal S}\phi=0$ is conformally invariant, 
since any solution is mapped onto another solution in the transformed
coordinates.\footnote{This concept of a dynamical symmetry, for the free diffusion equation, goes back to Jacobi (1842) and Lie (1881)
and was re-introduced into physics by Niederer (1972) \cite{Nied72}.}  A two-point function of quasi-primary scaling operators is 
$\mathscr{C}(t_1,t_2;r_1,r_2) := \langle \phi_1(z_1,\bar{z}_1) \phi_2(z_2,\bar{z}_2)\rangle = \langle \vph_1(t_1,r_1) \vph_2(t_2,r_2)\rangle$. 
Its covariance under ortho-conformal transformations is expressed by the `projective Ward identities' $X_n \mathscr{C}=Y_n \mathscr{C}=0$ for
$n=\pm 1,0$ \cite{Bela84}. For scalars, such that $\Delta_i=\overline{\Delta}_i=x_i$, this gives \cite{Poly70}
\BEQ \label{Cortho}
\mathscr{C}(t_1,t_2;r_1,r_2) = {\cal C}_0\, \delta_{x_1,x_2} \left( (t_1-t_2)^2 + (r_1-r_2)^2 \right)^{-x_1}
\EEQ
where ${\cal C}_0$ is a normalisation constant. 

An example of meta-conformal transformations in $(1+1)$ dimensions is given by \cite{Henk02}
\BEA
X_n   &=& -t^{n+1}\partial_t-\mu^{-1}[(t+\mu r)^{n+1}-t^{n+1}]\partial_r-(n+1)xt^n- (n+1)\frac{\gamma}{\mu}[(t+\mu r)^{n}-t^{n}]\nonumber\\
Y_{n} &=& -(t+\mu r)^{n+1}\partial_r- (n+1)\gamma (t+\mu r)^{n}
\label{Tmeta1}
\EEA
where $x,\gamma$ are the scaling dimension and the `rapidity' of the scaling operator $\vph=\vph(t,r)$ on which these generators act and 
the constant $1/\mu$ has the dimensions of a velocity. The Lie algebra $\langle X_n, Y_n\rangle_{n=\pm 1,0}$ 
is isomorphic to $\mathfrak{conf}(2)$ \cite{Henk15b}. 
An invariant equation\footnote{See \cite{Stoi15} for extensions as dynamical symmetries of the $(1+1)D$ Vlassov equation, isomorphic to 
$\mathfrak{conf}(2)$.} 
is ${\cal S}\vph=(-\mu\partial_t + \partial_r)\vph=0$, provided only that $\gamma=\mu x$, since the
only non-vanishing commutators of the Lie algebra with $\cal S$ are $\left[{\cal S},X_0\right]\vph = -{\cal S}\vph$ and 
$\left[{\cal S},X_{1}\right]\vph = -2t{\cal S}\vph+2(\mu x-\gamma)\vph$. The covariant two-point function is \cite{Henk02,Henk16a}
\BEQ \label{Cmeta1}
\mathscr{C}(t_1,t_2;r_1,r_2) = {\cal C}_0\: \delta_{x_1,x_2} \delta_{\gamma_1,\gamma_2} \left( t_1-t_2\right)^{-2x_1} 
\left( 1 + \frac{\mu}{\gamma_1} \left| \gamma_1 \frac{r_1-r_2}{t_1-t_2}\right| \right)^{-2\gamma_1/\mu}
\EEQ  
These well-known results are summarised in the first two columns of table~\ref{tab1}. 
Comparing the two-point functions (\ref{Cortho}) and (\ref{Cmeta1}) shows that even for the same
dynamical exponent $z=1$, different forms of the scaling functions are possible for ortho- and meta-conformal invariance. 

\begin{table}[tb]
\caption[tab1]{Comparison of ortho- and two examples of meta-conformal invariance. Listed are the commutators of the Lie algebra bases
$\langle X_n,Y_n\rangle_{n=\pm 1,0}\cong\mathfrak{conf}(2)$, the invariant Schr\"odinger operator $\cal S$ and the covariant two-point function
$\mathscr{C}(t;r)=\langle\vph(t,r)\vph(0,0)\rangle$, up to normalisation. 
The physical nature of $\mathscr{C}$ is also indicated.

For ortho-conformal invariance and meta-conformal invariance 1, one has the constraints $x_1=x_2$ and $\gamma_1=\gamma_2$. 
For the meta-conformal invariance 2, we list only case A from the text. One has $\mu^{-1}=\II \nu$ with $\nu>0$,
and the constraints $\gamma_1+\gamma_2=\mu$ and $\gamma_1-\gamma_2=\mu(x_1-x_2)$. \label{tab1}} 
\begin{center}\begin{tabular}{|c|lll|} \hline
        & \multicolumn{1}{c}{ortho} & \multicolumn{1}{c}{meta-1} & \multicolumn{1}{c|}{meta-2} \\ \hline
Lie     & $\left[ X_n, X_m\right] = (n-m) X_{n+m}$ & $\left[ X_n, X_m\right] = (n-m) X_{n+m}$    & $\left[ X_n, X_m\right] = (n-m) X_{n+m}$ \\
algebra & $\left[ X_n, Y_m\right] \:=(n-m)Y_{n+m}$ & $\left[ X_n, Y_m\right] \:=(n-m)Y_{n+m}$    & $\left[ X_n, Y_m\right] \:=(n-m)Y_{n+m}$ \\
$\mathfrak{conf}(2)$        
        & $\left[ Y_n,Y_m\right]\:\:=(n-m)X_{n+m}$ & $\left[Y_n, Y_m\right]\:\:=\mu(n-m)Y_{n+m}$ & $\left[ Y_n, Y_m\right] \:\:=\mu (n-m)Y_{n+m}$ 
        																													\\[0.14truecm] \hline
${\cal S}$    & $\partial_t^2 + \partial_r^2$      & $-\mu\partial_t + \partial_r$               & $-\mu\partial_t + \nabla_r$ \\[0.14truecm] \hline
$\mathscr{C}$ & $\left( t^2 + r^2\right)^{-x_1}$   & $t^{-2x_1} \left( 1 + \frac{\mu}{\gamma_1}\left|\frac{\gamma_1 r}{t}\right|\right)^{-2\gamma_1/\mu}$ 
                                                   & $t^{1-x_1-x_2}\cdot \nu t\left(\nu^2 t^2 + r^2\right)^{-1}$ \\[0.16truecm]
        & correlator                               & correlator                                  & response \hfill (case A)~~ \\ \hline      
\end{tabular}\end{center}
\end{table}

{\bf 4.} Are these examples of ortho- or meta-conformal invariance, which have $z=1$ and are realised in terms 
of local first-order differential operators, suitable as a dynamical symmetry of the {\sc dle} in $1+1$ dimensions~? 
This must be answered in the negative, for the following reasons. 
\begin{enumerate}
\item The {\sc dle} response function (\ref{7R}) is distinct from the predictions (\ref{Cortho},\ref{Cmeta1}), see also table~\ref{tab1}. 
For the meta-conformal two-point function (\ref{Cmeta1}), the functional form
is clearly different for finite values of the scaling variable $v=(r_1-r_2)/(t_1-t_2)$. The ortho-conformal two-point function (\ref{Cortho})
looks to be much closer, with the choice $x_1=\demi$ and the scale factor fixed to $\nu=1$, were it not for the extra factor $\nu(t-s)$. 
On the other hand, the two-time {\sc dle} correlator (\ref{7C}) does not agree with (\ref{Cortho}) either, but might be similar to a two-point function
computed in a semi-infinite space $t\geq 0$, $r\in\mathbb{R}$ with a boundary at $t=0$. 
\item The invariant equations ${\cal S}\vph=0$ are distinct from the deterministic part of the {\sc dle} Langevin equation (\ref{4}). 
Recall the well-known fact \cite{Pico04} that for Langevin equations ${\cal S}\vph=\eta$, where $\eta$ is a white noise, and where the noise-less equation
${\cal S}\vph_0=0$ has a local scale-invariance (including a generalised Galilei-invariance to derive  Bargman super-selection rules \cite{Barg54}) 
all correlators and response functions can be reduced to responses found in the noise-less theory. In particular, the two-time response
function of the full noisy equation $R(t,s;\vec{r})=R_0(t,s;\vec{r})$, is identical to 
the response $R_0$ found when the noise is turned off and computed from the dynamical symmetry \cite{Pico04,Henk10}.  

Indeed, in the example (\ref{6R},\ref{7R}) of the {\sc dle}, one sees that the two-time response $R$ is independent of $T$, 
which characterises the white noise. 
\end{enumerate}
We shall look for dynamical symmetries of the equation ${\cal S}\vph=\left(-\mu\partial_t + \nabla_r\right)\vph=0$, which is
the deterministic part of the 
{\sc dle} Langevin equation (\ref{4}), in $1+1$ dimensions. We shall seek to derive the form of the two-time response function 
$R(t,s;\vec{r})$ from this dynamical symmetry. 
The two-time correlator $C$ cannot be  obtained in this way. Rather, we shall see that its `deterministic' contribution
$C_0(t,s;\vec{r})=0$ simply vanishes. As shown in \cite{Pico04}, the correlator must be obtained from an integral over three-point response functions. 
We leave this for future work.  

{\bf 5.} In direct space, the invariant Schr\"odinger operator for {\sc dle} should be ${\cal S} := -\mu\partial_t + \nabla_r$, 
where $\nabla_r^{\alpha}$ denotes the Riesz-Feller fractional derivative \cite{Samk93} of order $\alpha$. For functions $f(r)$ of 
a single variable $r\in\mathbb{R}$ (assuming that $f(r)$ is such that the integral exists), we use the convention
\BEQ
\nabla_r^{\alpha} f(r) := \frac{\II^{\alpha}}{2\pi} \int_{\mathbb{R}^2} \!\D k\D x\: |k|^{\alpha}\: e^{\II k(r-x)}\, f(x)
\EEQ
Then the following properties hold true, for formal manipulations \cite{Baum07},\cite[app. J.2]{Henk10}
\BEA
\nabla_r^{\alpha}\nabla_r^{\beta} f(r) = \nabla_r^{\alpha+\beta} f(r) \;\; &,& \;\;
\left[ \nabla_r^{\alpha}, r \right] f(r) = \alpha \partial_r \nabla_r^{\alpha-2} f(r) \;\; , \;\; 
\nabla_r^{\alpha} f(\mu r) = |\mu|^{\alpha} \nabla_{\mu r}^{\alpha} f(\mu r) \nonumber \\
\nabla_r^{\alpha} e^{\II q r} = \left(\II |q|\right)^{\alpha} e^{\II q r} \;\; &,& \;\; 
\left(\wht{\nabla_r^{\alpha} f(r)}\right)(q) = \left(\II |q|\right)^{\alpha} \wht{f}(q) \;\; , \;\;
\nabla_r^2 f(r) = \partial_r^2 f(r)
\label{16}
\EEA
where $\wht{f}(q)$ is the Fourier transform of $f(r)$. In selecting the generators for the Lie algebra of dynamical symmetries, we follow \cite{Henk02}
and require that time translations $X_{-1}=-\partial_t$, dilatations $X_0 = -t\partial_t - r\partial_r -x$ and space translations $Y_{-1}$ are present. 
However, if one begins with the standard local generator $-\partial_r$ of spatial translations, it turns out that the {\it non-local} generator
$-\nabla_r$ is generated as well \cite{Baum07},\cite[ch. 5.3]{Henk10}. 
The closure of this set of generators, for generic values of $z\ne 2$, is still an open problem. 
In order to obtain a well-defined Lie algebra of dynamical symmetries of $\cal S$, we consider a {\it non-local} spatial translation operator
$Y_{-1}=-\nabla_r$. Consider the following set of single-particle generators
\BEA
X_{-1} &=& \!\!-\partial_t \;\; , \;\; X_0 = -t \partial_t - r \partial_r -x 
\;\; , \;\;
X_1 = -t^2\partial_t -2tr\partial_r -\mu r^2\nabla_r -2xt -2\gamma r\partial_r \nabla_r^{-1}
\label{1Part} 
\\
Y_{-1} &=& \!\!-\nabla_r \; , \;\; Y_0 = -t\nabla_r -\mu r \partial_r -\gamma 
\; , \; 
Y_1 = -t^2\nabla_r -2\mu tr \partial_r -\mu^2 r^2 \nabla_r -2\gamma t -2\gamma \mu r \partial_r \nabla_r^{-1}
\nonumber 
\EEA
As in the set (\ref{Tmeta1}) of meta-conformal transformations, the constants $x$ and $\gamma$, respectively, are the scaling dimension and rapidity of
the scaling operator $\vph=\vph(t,r)$ on which these generators act. It is now an afternoon's exercise (before tea time) to check, 
with the help of (\ref{16}),\footnote{Use the identities $\left[ \nabla_r, r^2\right]=2r\partial_r\nabla_r^{-1}$, 
$\left[ r^2\nabla_r,r\partial_r\right]=-r^2\nabla_r$ and $\left[ \nabla_r, \partial_r \right]=\left[\partial_r \nabla_r^{-1},r\right]=0$.} 
the following commutator relations, for $n,m\in\{\pm 1,0\}$
\BEQ \label{metaconf}
\left[ X_n, X_m\right] =(n-m) X_{n+m} \;\; , \;\; \left[ X_n, Y_m\right] =(n-m) Y_{n+m} \;\; , \;\; \left[ Y_n, Y_m\right] =\mu (n-m) Y_{n+m}
\EEQ
This establishes the Lie algebra isomorphism $\langle X_n,Y_n\rangle_{n=\pm 1,0}\cong\mathfrak{conf}(2)$. Furthermore, since
\BEQ
\left[ {\cal S}, Y_n \right]\vph = \left[ {\cal S}, X_{-1} \right]\vph = 0 \;\; , \;\;
\left[ {\cal S}, X_0 \right]\vph = - {\cal S}\vph \;\; , \;\;
\left[ {\cal S}, X_1 \right]\vph = -2t{\cal S}\vph +2(\mu x-\gamma)\vph
\EEQ
{\em the infinitesimal transformations (\ref{1Part}) form a Lie algebra of meta-conformal dynamical symmetries} (of the deterministic part) {\em of the  
{\sc dle} equation (\ref{4}), if $\gamma=x\mu$.} In contrast to the generators (\ref{Tmeta1}), 
the generators (\ref{1Part}) are non-local and do not generate
simple local changes of the coordinates $(t,r)$. In spite of an attempt to interpret non-local infinitesimal generators as the transformation
of a distribution of coordinates \cite{Henk11}, finding a clear geometrical interpretation of the generators (\ref{1Part}) remains an open problem. 

{\bf 6.} We look for the covariant $n$-point functions. We expect \cite{Henk10} that these will correspond physically to response functions, 
i.e. the two-time response $R(t,s;r)=\langle \vph(t,r)\wit{\vph}(s,0)\rangle$, 
where $\wit{\vph}$ is the response operator conjugate to the scaling operator $\vph$, in the
context of Janssen-de Dominicis theory \cite{Taeu14}. In order to write down the $n$-body operators analogous to (\ref{1Part}), we must ascribe 
a `signature' $\vep=\pm 1$ to each scaling operator. We choose the convention that $\vep_i=+1$ for scaling operators $\vph_i$ and $\vep_i=-1$ for
response operators $\wit{\vph}_i$. Then 
\BEA
Y_{-1} \:=\: Y_{-1}^{[n]} &=& \sum_i \left[ -\vep_i \nabla_i \right] \hspace{0.6truecm},\hspace{1.0truecm} 
Y_0 \:=\: Y_0^{[n]} \:=\: \sum_i \left[ -\vep_i t_i \nabla_i -\mu r_i D_i -\gamma_i \right] \nonumber \\
Y_1 \:=\: Y_1^{[n]} &=& \sum_i \left[ -\vep_i t_i^2 \nabla_i -2\mu t_i r_i D_i -\mu^2 \vep_i r_i^2\nabla_i -2\gamma_i t_i -2\mu \gamma_i \vep_i r_i D_i \nabla_i^{-1}\right]
\nonumber \\
X_{-1} \:=\: X_{-1}^{[n]} &=& \sum_i \left[ -\partial_i\right] \hspace{1.0truecm},\hspace{1.0truecm} 
X_0  \:=\: X_0^{[n]} \:=\: \sum_i \left[ -t_i\partial_i - r_iD_i - x_i \right] 
\label{2Part} \\
X_1 \:=\: X_1^{[n]} &=& \sum_i \left[ -t_i^2\partial_i -2t_ir_i D_i -\mu \vep_i r_i^2 \nabla_i -2x_i t_i -2\gamma_i \vep_i r_i D_i \nabla_i^{-1}\right]
\nonumber 
\EEA
with the short-hands $\partial_i=\frac{\partial}{\partial t_i}$, $D_i = \frac{\partial}{\partial r_i}$ and $\nabla_i = \nabla_{r_i}$. 
It can be checked that the generators (\ref{2Part}) obey the meta-conformal Lie algebra (\ref{metaconf}). Now, for a $(n+m)$-point function
\BD
\mathscr{C}_{n,m}=\mathscr{C}_{n,m}(t_1,\ldots,t_{n+m};r_{1},\ldots,r_{n+m}) 
= \left\langle \vph_1(t_1,r_1)\cdots\vph_n(t_n,r_n) \wit{\vph}_{n+1}(t_{n+1},r_{n+1})\cdots\wit{\vph}_{n+m}(t_{n+m},r_{n+m})\right\rangle
\ED
of quasi-primary scaling and response operators, the covariance is expressed through the projective Ward identities
$X_{k}^{[n+m]} \mathscr{C}_{n,m}=Y_{k}^{[n+m]}\mathscr{C}_{n,m}=0$, for $k=\pm 1,0$. 

{\bf 7.} We apply this to the two-time {\it response function} $\mathscr{R}=\mathscr{R}(t_1,t_2;r_1,r_2)=\mathscr{C}_{1,1}(t_1,t_2;r_1,r_2)$. From
$X_{-1}\mathscr{R}=0$ it follows that $\mathscr{R}=\mathscr{R}(t;r_1,r_2)$, with $t=t_1-t_2$. On the other hand, the condition $Y_{-1}\mathscr{R}=0$
would lead in Fourier space to $\left(\vep_1 |q_1| +\vep_2 |q_2|\right)\wht{\mathscr{R}}(t;q_1,q_2)=0$. Because of the assigned signatures
$\vep_1=-\vep_2=1$, this equation can have a non-vanishing solution such that we can write $\mathscr{R}=F(t,r)$, with $r=r_1-r_2$. 
However, for a two-point {\it correlator} $\mathscr{C}_{2,0}$, with $\vep_1=\vep_2=1$, the Ward identity $Y_{-1}\mathscr{C}_{2,0}=0$ would simply
imply that ${\wht{\mathscr{C}}}_{2,0}=0$, and in agreement with the fact that the {\sc dle}-correlator (\ref{6C},\ref{7C}) vanishes as $T\to 0$. 
Standard calculations, see e.g. \cite{Henk10}, lead to the following set of conditions for the function $\mathscr{R}=F(t,r)$
\begin{subequations} \label{cov}
\begin{align}
\left[ -t\partial_t -r\partial_r -x_1-x_2 \right] F &= 0 \label{covX0} \\
\left[ -t\vep_1 \nabla_r -\mu r\partial_r -\gamma_1 - \gamma_2 \right] F &= 0 \label{covY0} \\
\left[ -t^2\partial_t -2tr\partial_r -\mu r^2 \vep_1 \nabla_r -2x_1 t -2\gamma_1 \vep_1 r \partial_r \nabla_r^{-1} \right] F &= 0 \label{covX1} \\
\left[ -t^2 \vep_1 \nabla_r -2\mu tr\partial_r -\mu^2 \vep_1 r^2 \nabla_r -2\gamma_1 t -2\mu\gamma_1\vep_1 r\partial_r \nabla_r^{-1} \right] F &= 0
\label{covY1} 
\end{align}
\end{subequations}
Eqs.~(\ref{covX1},\ref{covY1}) can be further simplified by combining them with (\ref{covX0},\ref{covY0}) 
and reduce to
\BEQ \label{22}
(x_1-x_2) \left( t +\mu \vep_1 r \partial_r \nabla_r^{-1} \right) F = 0 \;\; , \;\; 
\left( \left(\gamma_1 - \gamma_2\right) - \mu \left( x_1 - x_2\right) \right) t F =0
\EEQ
If $F$ does not contain a factor $\sim \delta(t)$, the second eq.~(\ref{22}) gives the constraint $\gamma_1-\gamma_2=\mu(x_1-x_2)$. 
Eq.~(\ref{covX0}) implies the  scaling form $F(t,r)=t^{-2x} f(v)$, with $v=r/t$ and $x=\demi(x_1+x_2)$. From (\ref{covY0},\ref{22}), 
the scaling function $f(v)$ must satisfy, with $\gamma=\demi(\gamma_1+\gamma_2)$
\BEQ \label{23}
\left( \vep_1 \nabla_v +\mu v\partial_v +2\gamma \right) f(v) = 0 \;\; \mbox{\rm ~and~ } \;\;
\nabla_v^{-1} \left[ (x_1-x_2) \left( \vep_1 \nabla_v +\mu v\partial_v + \mu \right)\right] f(v) = 0 
\EEQ
The two conditions in eq.~(\ref{23}) are compatible in two distinct cases:
\begin{enumerate}
\item[{\bf ~Case A}:] \textcolor{blue}{\underline{\textcolor{black}{$2\gamma=\mu$}}}. 
Then $\left( \vep_1 \nabla_v +\mu v\partial_v + \mu \right) f(v) = 0$ and $x_1\ne x_2$ is still  possible. 
\item[{\bf ~Case B}:] \textcolor{blue}{\underline{\textcolor{black}{$x_1=x_2$}}}. 
Then $\gamma_1=\gamma_2$ and $\left( \vep_1 \nabla_v +\mu v\partial_v +2\gamma \right) f(v) = 0$. 
\end{enumerate}
In Fourier space, eq.~(\ref{covY0}) gives $\left( \II\vep_1 |q|  -\mu q\partial_q +(2\gamma-\mu)\right) \wht{f}(q)=0$, which illustrates
the difference between cases A and B. It follows that 
$\wht{f}(q) = \wht{f}_0 q^{2\gamma/\mu-1}\exp\left(\II\vep_1 |q|/\mu\right)$, where $\wht{f}_0$ is a normalisation constant. 
Finally, comparison of the Schr\"odinger operator ${\cal S}=-\mu\partial_t+\nabla_r$ 
with the {\sc dle} equation (\ref{4}) shows that $\mu^{-1}=\II \nu$. 
Transforming back into direct space, we find
\BEQ \label{scalf}
f(v) = f_0 \times \left\{ 
\begin{array}{ll} 
\vep_1 \nu  \left( \nu^2 + v^2 \right)^{-1}                                             & \mbox{\rm ~~;~ case A} \\
\Re \left( e^{-\II\pi \psi/2} \left( \vep_1 \nu - \II v \right)^{-\psi-1} \right) & \mbox{\rm ~~;~ case B, 
                                                                                           with $\psi+1:=2\II\nu\gamma$} 
\end{array} \right.
\EEQ
A linear combination of these two solutions is a solution of the linear system (\ref{23}) as well. 

{\bf 8.} In particular, for case A, the final form of the two-time response function $\mathscr{R}$ 
becomes, with a normalisation constant $F_0$ and $x=\demi(x_1+x_2)$
\BEQ
\mathscr{R} = F(t,r) = F_0\, t^{1-2x} \frac{\vep_1 \nu t}{\nu^2 t^2 + r^2} \;\; \mbox{\rm ~~;~ with $t=t_1-t_2$, $r=r_1-r_2$  ~(case A)} 
\EEQ
If one takes $x=\demi$, and $\nu\in\mathbb{R}_+$, this {\em reproduces the exact solution (\ref{7R}) of the response in $(1+1)D$ {\sc dle}}. 
This is our main result: {\em the non-local representation (\ref{2Part}) of $\mathfrak{conf}(2)$ is necessary to reproduce the correct 
scaling behaviour of the non-stationary response.} 
The properties and predictions of this second example of a meta-conformal symmetry, for the special case A, 
are listed in the last column of table~\ref{tab1}. An important difference is that ortho-conformal invariance
and meta-conformal invariance 1 predict the form of a two-time correlator $\mathscr{C}=\mathscr{C}_{2,0}$, 
whereas the meta-conformal invariance 2 predicts the form of a two-time response $\mathscr{R}=\mathscr{C}_{1,1}$.

\underline{Summary:} we have proposed a meta-conformal dynamical symmetry for the {\sc dle} in $1+1$ dimensions. This symmetry, isomorphic to the
Lie algebra $\mathfrak{sl}(2,\mathbb{R})\oplus\mathfrak{sl}(2,\mathbb{R})$, is realised in terms of {\em non-local} generators, see 
eqs.~(\ref{1Part},\ref{2Part}). It is distinct from other known representations of the conformal Lie algebra, both in the form of
the invariant Schr\"odinger operator $\cal S$ and in the predicted shape of the covariant two-point function, see table~\ref{tab1}. 
In particular, the generator $Y_{-1}$ which plays the role of `spatial translations' is manifestly non-local. 
The full time-space form of the two-time response $R(t,s;r)$ in (\ref{7R}) can be derived from this dynamical symmetry. This is the first
time that (i) the full response (and not only the auto-response $R(t,s;0)$) can be confirmed and (ii) that the set of generators closes 
into a Lie algebra, for a system with $z\ne 2$. 

In view of Spohn's mapping \cite{Spoh99}, which relates the $(1+1)D$ {\sc dle} equation (\ref{4}) with the quantum chain (\ref{xxz})
and the terrace-step-kink model, a convenient linear combination of the prediction of $R(t,s;r)$ from cases A and B might 
describe the non-stationary response of vicinal surfaces. Indeed, the explicit form of the connected {\em stationary} 
correlator $\left\langle\vro(t,r)\vro(t,0)\right\rangle_c$ of the particle density $\vro(t,r)=\partial_r h(t,r)$, obtained
by Karevksi and  Sch\"utz from (\ref{xxz}) in the limit $\lambda\to\infty$ \cite[eq.(1)]{Kare16}, 
contains two terms which look quite analogous to the responses (\ref{scalf}) in cases A and B. 
While that is distinct from the non-stationary responses considered here, the qualitative analogy is encouraging.
Certainly, a precise test is called for. This will require to work out higher $n$-point
functions in order to be able to derive the form of non-equilibrium correlators. Stationary correlators might be included by considering
an appropriate initial condition. An obvious further extension will be to dimensions $d>1$. 
Conceptually, the consideration of
manifestly non-local generators in local scale-invariance 
might lead to further insight for the construction of dynamical symmetries for different values of $z$. 


\noindent 
{\bf Acknowledgements:}  
I thank G.M. Sch\"utz for fruitful discussions. 
This work was started at the workshop `Advanced Conformal Field Theory and Applications' ({\sc acft}) at the Institute Henri Poincar\'e Paris. 
It is a pleasure to thank the organisers for their warm hospitality. This work was also partly supported by the 
Coll\`ege Doctoral franco-allemand Nancy-Leipzig-Coventry ({\it `Statistical Physics of Complex Systems'}) of UFA-DFH. 


\end{document}